\documentclass[aps,twocolumn,superscriptaddress]{revtex4}

\usepackage{revsymb,graphicx,bbm}

\usepackage[english]{babel}

\begin{document}

\newcommand{\bra}[1]{\langle #1|}
\newcommand{\ket}[1]{|#1\rangle}
\newcommand{\braket}[2]{\langle #1|#2\rangle}

\title{Noiseless linear amplification via weak measurements}

\author{David Menzies}
\email{djm12@st-andrews.ac.uk} \affiliation{School of Physics and Astronomy, University of St. Andrews, North Haugh, St. Andrews KY16 9SS, UK.}
\date{\today}

\author{Sarah Croke}
\email{scroke@perimeterinstitute.ca} \affiliation{Perimeter Institute for Theoretical Physics, 31 Caroline St N, Waterloo, Ontario N2L 2Y5, Canada.}
\date{\today}


\begin{abstract}
We discuss the recently introduced concept of non-deterministic noiseless linear amplification, demonstrating that such an operation can only be performed perfectly with vanishing probability of success.  We show that a weak measurement, which imprints the weak value of an observable of a pre-selected and post-selected system onto a probe system, can be used to approximate probabilistic noiseless amplification.  This result may be applied to various tasks in continuous variable quantum information, including entanglement concentration, probabilistic cloning, and in quantum repeaters.  We discuss the application of our scheme to probabilistic cloning of weak coherent states.
\end{abstract}

\maketitle

When a pre and post-selected quantum system interacts weakly with a probe system, the so-called weak value of an observable of the system is imprinted onto the probe \cite{Aharonov88,Aharonov90}.  The weak value can take values outside the range of the eigenvalues of the associated observable \cite{Aharonov88}, and can even be complex \cite{Jozsa07}.  Weak measurements have been studied extensively to investigate the properties of quantum systems in between measurements, in the time symmetric or two-state vector formalism \cite{Aharonov64,Reznik95}, and have proven to be a fruitful concept in studying paradoxes in quantum mechanics \cite{Aharonov05b}.  Aside from their theoretical investigation, weak values have also been experimentally observed in a number of quantum optical systems \cite{Pryde05,Ralph06b,Agarwal07}.  Taking a more operational approach, the imprinting of weak values can be exploited to perform useful tasks in quantum information theory, such as in quantum communication schemes \cite{Botero00}, in entanglement concentration \cite{Menzies07,Menzies08}, and to amplify small optical effects, allowing them to be measured \cite{Hosten08,Resch08}.  Nevertheless, this potentially useful effect has remained relatively unexplored.  In this paper we will discuss the use of the weak measurement formalism to probabilistically perform approximate noiseless linear amplification on small amplitude quantum optical states.  Although deterministic noiseless linear amplification is forbidden by the no-cloning theorem \cite{noclone}, it was recently pointed out that this does not rule out performing the operation non-deterministically \cite{Ralph08}, as long as the distinguishability of non-orthogonal states on average does not increase.  We consider explicitly the operation that performs noiseless amplification, and show that, even in principle, it can only occur with vanishing probability of success, and therefore cannot be implemented exactly.  However, it can be approximated arbitrarily well, with the probability of success necessarily decreasing as the approximation gets better.  We present a weak measurement framework for constructing protocols to achieve this.  Such protocols may be expected to have many uses in quantum information and communications including, for example, entanglement concentration \cite{Menzies07,Menzies08}, probabilistic cloning \cite{Fiurasek04}, and in quantum repeaters \cite{vanLoock08}.  We apply our result to the probabilistic cloning of small amplitude coherent states and compare the performance of our scheme to the linear optical proposal of Ralph and Lund \cite{Ralph08}.

Linear or phase-insensitive amplification of the complex amplitude of a field mode is associated with a transformation of the annihilation operator $\hat{a} \rightarrow g \hat{a}$ for some $g > 1$.  Clearly there can be no unitary operator achieving such a transformation, since the commutator $[ \hat{a},\hat{a}^{\dagger} ]$ is not preserved.  The usual interpretation is that amplification must add noise to preserve the commutator, and the minimum amount of noise that must be added is well-studied \cite{Caves82}.  Alternatively, we can ask if there is a non-unitary operator achieving this transformation probabilistically \cite{Ralph08}.  It is useful to consider amplification in the Schr\"{o}dinger picture, in which it can be thought of as the operation that transforms a coherent state $\ket{\alpha}$, an eigenvector of the annihilation operator with eigenvalue $\alpha$, to $\ket{g \alpha}$, with some probability of success $p_{S}$, which may depend on the input state.  Thus, we define the operator $\hat{\Gamma} (g)$, corresponding to noiseless amplification, by its action on coherent states
\begin{equation}
\hat{\Gamma}(g) \ket{\alpha} = \sqrt{p_{S}} \ket{g \alpha}.
\end{equation}
In the photon number basis this may be written
\begin{equation}
\hat{\Gamma}(g) e^{-|\alpha|^2/2} \sum_{n=0}^{\infty} \frac{\alpha^n}{\sqrt{n !}} \ket{n} = \sqrt{p_{S}} e^{- |g \alpha|^2/2} \sum_{n=0}^{\infty} \frac{(g \alpha)^n}{\sqrt{n!}} \ket{n}.
\end{equation}
Thus $\hat{\Gamma} (g)$ has the form
\begin{equation}
\hat{\Gamma}(g) = c \sum_{n=0}^{\infty} g^n \ket{n} \bra{n} = c g^{\hat{n}}
\label{Gamma}
\end{equation}
where $\hat{n} = \sum n \ket{n} \bra{n}$ is the photon number operator, and the constant of proportionality $c$ determines the probability of success via $p_{S} = |c|^2 e^{-(g^2 -1) |\alpha|^2}$.  We now note that any physically-allowed operation is described mathematically by a trace non-increasing completely positive map, and its action on some arbitrary input state $\hat{\rho}$ has a Kraus operator decomposition \cite{Nielsen00}
\begin{equation}
\hat{\rho} \rightarrow \sum_i \hat{A}_i \hat{\rho} \hat{A}_i^{\dagger},
\end{equation}
where the Kraus operators $\{ \hat{A}_i \}$ satisfy the constraint $\sum_i \hat{A}_i^{\dagger} \hat{A}_i \leq \hat{{1}}$ (with equality if and only if the operation is deterministic).  In our case there is only one Kraus operator $\hat{A}_1 = \hat{\Gamma}(g)$, and thus we require $\hat{\Gamma}^{\dagger}(g) \hat{\Gamma}(g) = |c|^2 g^{2 \hat{n}} \leq \hat{{1}}$.  The constraint is therefore
\begin{equation}
|c|^2 \sum_{n=0}^{\infty} g^{2n} \ket{n} \bra{n} \leq \sum_{n=0}^{\infty} \ket{n} \bra{n},
\end{equation}
which, for $g>1$ can only hold if $|c|^2 = 0$, i.e. the operation has vanishing probability of success.  Despite this, we note that noiseless amplification can be approximated arbitrarily well with finite probability.  One way to achieve this in principle \cite{Fiurasek04} is by truncating the Hilbert space, and performing the operation
\begin{equation}
\hat{\Gamma}_{approx}(g) = c_N \sum_{n=0}^{N} g^{n} \ket{n} \bra{n}
\end{equation}
on the subspace spanned by the first $N+1$ Fock states, for some arbitrarily large but finite $N$.  This is an allowed operation provided $|c_N|^2 g^{2N} \leq 1$, thus as the approximation gets better the probability of success, which is proportional to $|c_N|^2$, falls off exponentially.

In the weak measurement formalism, a system prepared in some state $\ket{\phi_i}$ is allowed to interact weakly with another, probe system, through an interaction Hamiltonian of the form $\hat H_I = \hbar \kappa(t) \hat A \otimes \hat P$, and is later post-selected in state $\ket{\phi_f}$.  If the interaction persists for a time $T$, we denote $\kappa_T = \int_0^{T} \kappa(t) {\rm d} t$, and a probe system initially in state $\ket{\psi_i}$ is transformed via
\begin{equation}
\ket{\psi_i} \rightarrow \bra{\phi_f} e^{-i \kappa_T \hat A \otimes \hat P} \ket{\phi_i} \ket{\psi_i} \approx \braket{\phi_f}{\phi_i} e^{-i \kappa_T A_W \hat P} \ket{\psi_i},
\label{weakmeas}
\end{equation}
where we have neglected normalisation and
\begin{equation}
A_W = \frac{\bra{\phi_f} \hat A \ket{\phi_i}}{\braket{\phi_f}{\phi_i}}.
\end{equation}
is the weak value of the observable $\hat A$.  A weak measurement is defined by the approximation in eqn (\ref{weakmeas}), which is valid if the coupling is weak enough and for certain probe states.  Thus the probe system registers the weak value of the observable $\hat A$, and not one of its eigenvalues.  Note that the weak value $A_W$ is complex in general \cite{Jozsa07}; the real part gives an effective unitary evolution for the probe system, while a non-zero imaginary part results in non-unitary evolution.

This paradigm suggests a way to approximately implement probabilistic noiseless amplification on the probe state.  If we perform a weak measurement with an interaction Hamiltonian of the form $\hat H_I = \hbar \kappa(t) {\hat O} \otimes {\hat n}$, the probe system is transformed via
\begin{equation}
\ket{\psi_f} \propto  e^{-i \kappa_T \Re (O_W) \hat n} (e^{\kappa_T \Im (O_W)})^{\hat n} \ket{\psi_i},
\label{weakterm}
\end{equation}
where $O_W = \Re(O_W) + i \Im(O_W)$ is the weak value of $\hat{O}$, the actual value of which depends on the pre- and post-selections.  Identifying $g=e^{\kappa_T \Im(O_W)}$, it is clear that the non-unitary part approximates noiseless amplification (eqn \ref{Gamma}) when $\Im(O_W) > 0$.  Note that the real part of the weak value simply imparts a known phase shift, given by $\kappa_{T} \Re (O_W(\omega))$, which can easily be corrected if necessary by adjusting the optical path length.
\begin{figure}[!h]
\centering
\includegraphics[height=15mm]{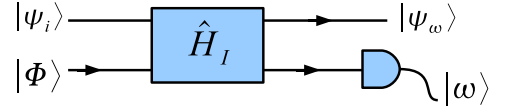}
\caption{Weak measurement model of noiseless amplification.  The weak value of the pre and post-selected system is imprinted on the probe system.  For an interaction Hamiltonian of the form $\hat H_I = \hbar \kappa(t) {\hat O} \otimes {\hat n}$, and certain post-selections, the transformation of the probe approximates noiseless amplification.} \label{probamp}
\end{figure}
Our weak measurement model for noiseless amplification is thus based on the configuration depicted in Fig.\ref{probamp}; the probe state in mode $A$ is coupled to an ancilla state in mode $B$ prepared in $\ket{\Phi}$, by means of unitary evolution under a Hamiltonian $\hat H_I = \hbar \kappa(t) {\hat O} \otimes {\hat n}$.  The ancilla is later post-selected in state $\ket{\omega}$, by means of a measurement on mode $B$.  The transformed state of the probe system, denoted $\ket{\psi_\omega}$, depends on the measurement outcome, and is given by
\begin{eqnarray}
\ket{\psi_\omega} &=& {\cal N} \bra{\omega} e^{-i \kappa_T \hat O \hat n} \ket{\Phi} \ket{\psi_i} \nonumber \\
&\approx& {\cal N} \bra{\omega} (\hat{{1}} -i \kappa_T \hat O \hat n) \ket{\Phi} \ket{\psi_i} \\
&\approx& {\cal N} \braket{\omega}{\Phi} e^{-i \kappa_T O_W(\omega) \hat n} \ket{\psi_i} \nonumber
\end{eqnarray}
where ${\cal N}$ is a normalisation constant, and the notation $O_W (\omega)$ reflects the fact that the weak value depends on the result of measurement of mode $B$.  Expanding the actual output state and the approximated state in the eigenbasis of the number operator $\hat n$ gives the infinite set of equations
\begin{equation}
\left( \frac{\bra{\omega} e^{-i \kappa_T \hat O n} \ket{\Phi}}{\braket{\omega}{\Phi}} - e^{-i \kappa_T O_W(\omega) n} \right) \braket{n}{\psi_i} \approx 0, \label{weakness}
\end{equation}
$\forall n \in [0, \infty )$.  These weakness conditions must be satisfied if the weak measurement model is to be considered an accurate description of this indirect measurement.  They are automatically satisfied for small $n$ provided that the coupling between signal and probe is weak $\kappa_T \ll 1$, and for large or intermediate $n$ can be satisfied provided $\braket{n}{\psi_i} \approx 0$.  States which have this form include, for example, small amplitude coherent states, and small amplitude squeezed states.

According to eqn (\ref{weakterm}), to achieve noiseless amplification, we need to choose a configuration that allows weak values with a positive imaginary part.  Consider that an expectation value can be written as a probabilistic mixture of weak values \cite{Aharonov05} $ \langle \Phi \vert {\hat O} \vert \Phi \rangle = \sum_{\omega \in \Omega} \vert \langle \omega \vert \Phi \rangle \vert^2 O_W(\omega)$, and so $\Im (\langle \Phi \vert {\hat O} \vert \Phi \rangle) = \sum_{\omega \in \Omega} \vert \langle \omega \vert \Phi \rangle \vert^2 \Im \{ O_W(\omega) \}$. However, since all expectation values for any observable are necessarily real then $\sum_{\omega \in \Omega} \vert \langle \omega \vert \Phi \rangle \vert^2 \Im \{ O_W(\omega) \} = 0$. Thus, there are two possibilities; the trivial case $\Im \{ O_W(\omega) \} = 0$ for all $\omega \in \Omega$, and the case $\exists \omega \in \Omega^+$ such that $\Im \{ O_W(\omega) \} > 0$ and $\exists \omega \in \Omega^-$ such that $\Im \{ O_W(\omega) \} < 0$.  Thus this procedure is probabilistic since only a subset of all measurement outcomes $\omega$ accompany a weak value such that $\Im \{ O_W(\omega) \} > 0$.  The measurement on mode $B$ may be described by a probability operator measure (POM) \cite{Helstrom76}, also known as a positive-operator valued measure (POVM) \cite{Peres93}, that is a set of positive operators $\hat{\Pi}_{\omega}$ satisfying $\sum_{\omega} \hat{\Pi}_{\omega} = \hat{1}$ in the discrete case, or $\int {\rm d} \omega \, \hat{\Pi}_{\omega} = \hat{1}$ in the continuous case.  Post-selecting state $\ket{\omega}$ is associated with a POM element of the form $\hat{\Pi}_{\omega} = a_{\omega} \ket{\omega} \bra{\omega}$ for some positive constant $a_{\omega}$, and the probability of obtaining the corresponding measurement outcome may be expressed
\begin{eqnarray}
P(\omega) &=& {\rm Tr} \left( e^{- i \kappa_T {\hat n} {\hat O}} \ket{\Phi} \ket{\psi_i} \bra{\psi_i} \bra{\Phi} e^{i \kappa_T {\hat n} {\hat O}} (\hat{\Pi}_{\omega} \otimes \hat{1}_p) \right) \nonumber \\
&=& a_{\omega} {\rm Tr} \left( \bra{\omega} e^{- i \kappa_T {\hat n} {\hat O}} \ket{\Phi} \ket{\psi_i} \bra{\psi_i} \bra{\Phi} e^{i \kappa_T {\hat n} {\hat O}} \ket{\omega} \right) \nonumber \\
&\approx& a_{\omega} |\braket{\omega}{\Phi}|^2 \langle \psi_i \vert e^{2 \kappa_T \Im \{ O_W(\omega) \} {\hat n}} \vert \psi_i \rangle
\label{prob}
\end{eqnarray}
where $\hat{1}_p$ is the identity operator on the probe system.  Thus the probability of success of the noiseless amplification is $P_{S} = \sum_{\omega \in \Omega^+} P(\omega)$.  An analogous expression may be obtained in the continuous POM case, where the probability $P(\omega)$ in eqn (\ref{prob}) becomes a probability density $\rho(\omega) {\rm d} \omega$, and the probability of success is given by $\int_{\Omega^+} \rho(\omega) {\rm d} \omega$.  Note that, despite our earlier discussion, this probability need not be vanishing as this scheme is only an approximation to noiseless amplification.  Further, our scheme is only applicable to states which satisfy the weakness conditions, that is those with negligible support at large $n$.  Note that these are precisely the states for which the theoretical method of truncating the state space method provides a good approximation to noiseless amplification.  In the protocols discussed below, we propose to measure the accuracy of our approximation via the fidelity ${\cal F} = \vert \langle \psi_{\omega}^{(G)} \vert \psi_{\omega} \rangle \vert^2$ between the final probe state in the weak approximation $\ket{\psi_{\omega}}$, as defined in (\ref{weakterm}), and the final probe state given by the general quantum-mechanical formula, $\ket{\psi_{\omega}^{(G)}} = {\cal N} \bra{\omega} e^{- i \kappa_T {\hat O} {\hat n}} \ket{\Phi} \ket{\psi_i}$.  The better the approximation provided by the weak measurement model to the actual output of the apparatus, the higher the fidelity.

An advantage of this weak measurement approach is that many different physical apparatus may be described by the same formalism.  We require an interaction Hamiltonian of the form $\hat{H}_I = \hbar \kappa(t) \hat{O} \otimes \hat{n}$, along with the ability to prepare and measure quantum states.  In principle a wide range of combinations of physical ancilla systems, interactions and measurements can give rise to a physical realisation of approximate noiseless amplification.  This has already been shown to be a powerful approach in demonstrating that existing models of Gaussian continuous variable entanglement concentration can all be thought of as special cases of the weak measurement formalism considered here \cite{Menzies07}.
\begin{figure}[!h]
\centering
\includegraphics[height=55mm]{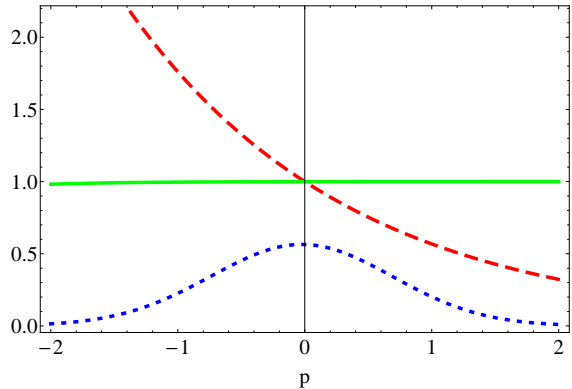}
\caption{Probabilistic cloning of weak coherent states. Graph shows the probability density $\rho(p)$ (blue dotted line), the gain $g(p)$ (red dashed line) and the fidelity $F(p)$ between the target state and the weak approximation (green solid line), for parameter values $\kappa_T = 4 \times 10^{-5}$, $\alpha = 10^4$ and $\beta = 0.2$.}
\label{cloneweak}
\end{figure}
We consider now the application of this scheme to cloning of weak coherent states, and discuss the performance of one possible physical implementation.  The configuration we consider is an all optical setting where an ancilla initially prepared in a coherent state $\ket{\alpha \in \Re}$ interacts with the probe system through the cross Kerr effect ${\hat H}_I = \hbar \kappa {\hat n}_A {\hat n}_B$, and the measurement corresponds to balanced homodyne detection, post-selecting onto the quadrature eigenstate $\ket{p}$.  Thus the weak value imprinted on the probe is that of the number operator for the ancilla mode, and is given by $n_W = \frac{\bra{p} \hat{n} \ket{\alpha}}{\braket{p}{\alpha}} = - \alpha^2 - i \sqrt{2} \alpha p$.  Cloning of a small amplitude coherent state $\ket{\beta}$ is equivalent to amplification $\ket{\beta} \rightarrow \ket{\sqrt{2} \beta}$, since from this a 50-50 beam-splitter can produce two copies of $\ket{\beta}$.  In fact for any $g \geq \sqrt{2}$ it is possible to extract two copies of $\ket{\beta}$ from $\ket{g \beta}$ using linear optics.  Thus our success condition may be expressed $g = e^{\kappa_T \Im \{O_W(\omega) \}} \geq \sqrt{2}$.  This in turn requires $p \leq - \frac{ \ln 2}{2 \sqrt{2} \alpha \kappa_T}$.  The POM associated with the quadrature measurement is a continuous POM, with operators $\{ \hat{\Pi}_p {\rm d}p = \ket{p} \bra{p} {\rm d} p \}$, and the probability density may be calculated to be $\rho(p) = (e^{-p^2 + (g^2-1) \beta^2})/\sqrt{\pi}$.  In Fig. \ref{cloneweak} we plot the probability density $\rho(p)$, the gain $g = e^{- \sqrt{2} \alpha p \kappa_T }$ and the fidelity ${\cal F}(p) = \vert \langle \psi_f^{(G)}(p) \vert \psi_f(p) \rangle \vert^2$ for physically realistic parameter values of $\kappa_T = 4 \times 10^{-5}$, $\alpha = 10^4$ and $\beta = 0.2$.  We note that the fidelity decreases as the gain increases, and we choose the range of measurement outcomes for which $-1.6 \leq p \leq - \frac{ \ln 2}{2 \sqrt{2} \alpha \kappa_T}$ to be those for which cloning is considered to be successful.  This corresponds to $g \geq \sqrt{2}$, but also ensures high fidelity of cloning.  In Table \ref{compare} we give some numerical values for the fidelities and corresponding probabilities of success achievable by our model, and give those achieved by the linear optical model of \cite{Ralph08} for comparison purposes.  For fidelities comparable to those achieved by this model, our protocol has a probability of success that is an order of magnitude larger.  Further, at a fidelity of $0.99$ our protocol gives a probability of success as high as $20 \%$.
\begin{table}[h]
\begin{tabular}{c|c|c|c}
Model & $\beta$ & Fidelity & $P_S$ \\
\hline
Weak model, $\kappa_T = 4 \times 10^{-5}$ & 0.2 & $>$ 0.99 & 20 \% \\
Weak model, $\kappa_T = 2 \times 10^{-5}$ & 0.2 & $>$ 0.9996 & 4 \% \\
Weak model, $\kappa_T = 2 \times 10^{-5}$ & 0.5 & $>$ 0.995 & 4 \% \\
Linear optics & 0.2 & $>$ 0.999 & 0.5 \% \\
Linear optics & 0.5 & $\approx$ 0.99 & 0.5 \%
\end{tabular}
\caption{Table comparing the performance of our weak measurement protocol to Ralph and Lund's linear optics proposal.  For the weak measurement model in each case $\alpha = 10^4$, success corresponds to $-1.6 \leq p \leq - \frac{ \ln 2}{2 \sqrt{2} \alpha \kappa_T}$, and we specify the strength of the coupling $\kappa_T$.  Values for the linear optics model are obtained from the graphs in Fig 3 of \cite{Ralph08}.}
\label{compare}
\end{table}

We have considered noiseless linear amplification, and shown that it is an operation that can only be performed perfectly with vanishing probability of success.  Nevertheless, it can be approximated arbitrarily well, and we have introduced a weak measurement framework for constructing protocols to achieve this.  Our framework potentially describes a range of possible physical realisations, all that is required is a suitable interaction Hamiltonian, and the ability to prepare and measure states.  The resulting protocols achieve approximate noiseless amplification with some probability of success, and success is heralded.  Our scheme is of theoretical interest as an example of how weak values may be exploited to perform useful tasks in quantum information theory.  We have discussed one possible physical implementation, in an all optical setting, and have shown how this may be applied to probabilistically clone small amplitude coherent states.  This implementation is conceptually simple, requiring only a small cross-Kerr non-linearity, coherent state inputs, and balanced homodyne detection.  The effects of experimental imperfections, and in particular how the fully continuous time, multi-mode analysis of the cross-Kerr effect considered in \cite{Shapiro07,Leung08} affects our all-optical scheme are left for future work.  Further, our weak measurement framework allows the identification of working configurations, but leaves the question of optimization unanswered.  Investigating this problem and the wider applicability of weak measurements in quantum information is also left for future work.

{\it Acknowledgements}. This work was supported by the {\it Engineering and Physical Sciences Research Council}, the EU Grant No. FP7 212008COMPAS (DM) and by the Perimeter Institute for Theoretical Physics.  Research at Perimeter Institute is supported by the Government of Canada through Industry Canada and by the Province of Ontario through the Ministry of Research \& Innovation.

\end{document}